\documentclass[manuscript]{acmart}

\usepackage{multirow} 

\AtBeginDocument{%
  \providecommand\BibTeX{{%
    \normalfont B\kern-0.5em{\scshape i\kern-0.25em b}\kern-0.8em\TeX}}}

\setcopyright{acmcopyright}
\copyrightyear{2022}
\acmYear{2022}

\acmConference[Workshop on CARS with RecSys '22]{Workshop on Context-Aware Recommender Systems (CARS), in conjunction with 16th ACM Conference on Recommender Systems}{September 23,
  2022}{Seattle, WA}
\acmPrice{15.00}
\acmISBN{978-1-4503-XXXX-X/18/06}

\begin{document}

\title{The Long Tail of Context: Does it Exist and Matter?}

\author{Konstantin Bauman}
\email{kbauman@temple.edu}
\orcid{0000-0003-4964-9882}
\affiliation{%
  \institution{Fox School of Business, Temple University}
  \streetaddress{P.O. Box 1212}
  \city{Philadelphia}
  \state{Pennsylvania}
  \country{USA}
  \postcode{43017-6221}
}

\author{Alexey Vasilev}
\email{alexxl.vasilev@yandex.ru}
\affiliation{%
  \institution{Sber, AI Lab}
  \city{Moscow}
  \country{Russian Federation}
}

\author{Alexander Tuzhilin}
\email{kbauman@temple.edu}
\affiliation{%
  \institution{Stern School, New York University}
  \city{New York}
  \state{New York}
  \country{USA}
}

\renewcommand{\shortauthors}{Bauman, et al.}

\begin{abstract}
  Context has been an important topic in recommender systems over the past two decades. 
A standard representational approach to context assumes that contextual variables and their structures are known in an application. 
Most of the prior CARS papers following representational approach manually selected and considered only a few crucial contextual variables in an application, such as time, location, and company of a person. This prior work demonstrated significant recommendation performance improvements when various CARS-based methods have been deployed in numerous applications.
However, some recommender systems applications deal with a much bigger and broader types of contexts, and manually identifying and capturing a few contextual variables is not sufficient in such cases. 
In this paper, we study such ``context-rich'' applications dealing with a large variety of different types of contexts. We demonstrate that supporting only a few most important contextual variables, although useful, is not sufficient.  
In our study, we focus on the application that recommends various banking products to commercial customers within the context of dialogues initiated by customer service representatives. In this application, we managed to identify over two hundred types of contextual variables.
Sorting those variables by their importance forms the Long Tail of Context (LTC). 
In this paper, we empirically demonstrate that LTC matters and using all these contextual variables from the Long Tail leads to significant improvements in recommendation performance.

\end{abstract}

\ccsdesc[500]{Information systems~Collaborative filtering}

\keywords{context-aware recommender systems, long tail of context, dialogue-based recommendations}

\maketitle

\section{Introduction}
\label{sec:intro}


Context has been extensively studied in recommender systems over the last 20 years \citep{villegas2018characterizing, del2021ai} and numerous contributions have been made to the CARS field \cite{CARSChapter2022}.
A standard representational approach to CARS \citep{Dourish:2004, CARSChapter2022} assumes that all contextual variables, including their structure and values, are known a-priori in a given application.
Most of the prior CARS-related research following representational approach manually selected and considered only a few crucial contextual variables in an application, such as time, location, company of a person, 
that are either provided by the user \cite{Odic2013PredictingAD} or learned from the application data, such as hotel reviews \cite{HaririMobasherReviewMining2011}. This method worked really well in practice across different cases, such as shopping, travel, music, restaurants and other applications, where recommendation performance was significantly improved, as shown in \cite{Tuzhilin2005,Odic2013PredictingAD,HaririMobasherReviewMining2011,McInerneyRecSys2019,DragoneWWW2019}.

However, some other recommender systems, such as dialogue-based systems in customer service applications, deal with a much bigger and broader variety of contexts, and manual selection of only a few contextual variables is not sufficient in such cases. For example, consider the dialogue system of customer service representatives talking to the customers regarding products and services, and the opportunities and problems associated with them. Since there are several products and services and various types of customers dealing with a variety of issues, the set of contextual variables is significantly larger than in such applications as restaurant or movie recommendations.
For example, in the dialogue-based banking application considered in this paper, we identified over two hundred different types of context.

In this paper, we study such ``context-rich'' applications dealing with a large variety of different types of contexts. We demonstrate that supporting only a few most important contextual variables, although useful, is not sufficient because we loose a lot by omitting the remaining less important types of context that, collectively, is as crucial for recommendation performance as a few most important types contextual variables. 
As a part of this study, we focus on the application that recommends various banking products to commercial customers within the context of dialogues initiated by bank customer service representatives as a part of the Customer Relationship Management (CRM) process. 
We managed to identify over two hundred types of context in this application by adapting the methods of extracting contextual variables from user-generated reviews described in \cite{bauman2021know}. 
We show that this long list of contextual variables includes a large number of infrequent variables forming a Long Tail of Context (LTC).

We also empirically demonstrate in this paper that the Long Tail of Context matters in the sense that using all these contextual variables, including those in the Long Tail, leads to significantly better recommendations vis-\'a-vis the case of relying only on those from the Head of the distribution. For example, we observe that the recommendation performance improvement in terms of AUC measure ranges from 8\% to 15\% when comparing the use of all contextual variables vs. only 10\% of the most frequent ones.  In particular, we demonstrate this by analyzing the dataset of transcribed dialogues between call center managers and corporate clients of a large European Bank. We extract contextual variables mentioned by clients and then we train a model to predict customer's propensity to purchase products offered by managers using those variables. We analyze the power of LTC by using progressively more contextual variables to train our model and evaluate its recommendation performance. 

In this paper, we make the following contributions. We demonstrate that
\begin{enumerate}
    \item the \textit{Long Tail of Context (LTC) exists} in some recommendation applications, such as dialogue systems
    \item \textit{LTC matters} in the sense that using all the contextual variables leads to significantly better recommendation performance than when only a few most important variables from the Head are used.
\end{enumerate}


\section{Literature Review}
\label{sec:literature}


\subsection{Context-Aware Recommender Systems}\label{sec:cars}

Context is a complex and multifaceted concept that has been studied for many years \cite{CARSChapter2022}. In recommender systems, contextual information is defined by a group of variables that are independent of the user and item attributes, that reflect user's circumstances while consuming the items, and that affect user's preferences, such as time, location and weather \cite{CARSChapter2022}.
It has been already shown that contextual information can help to provide better recommendations in many different applications, such as movies \cite{Odic2013PredictingAD, Wu:2017:IPT:3103645.3103886}, music \cite{DSBDA2018, DragoneWWW2019, McInerneyRecSys2019, hansen2020contextual}, 
hotels \cite{HaririMobasherReviewMining2011, Chen:2015}, and restaurants \cite{Li:2010:CRB:1944566.1944645, 10.1007/978-3-319-08786-3_6}.

As mentioned in Section \ref{sec:intro}, most of the research on CARS assumes the representational view \citep{Dourish:2004, CARSChapter2022}, where all contextual variables, including their structure and values, are known a-priori in a given application. As opposed to traditional representational view, in the latent approach, contextual information is not observed and instead can be modeled using some machine-learning and deep-learning methods, e.g., \cite{Ding2019, xin2019cfm, unger2020context}. In this paper, we do not consider latent modeling approach as we focus on contextual variables that were \textit{explicitly} mentioned by bank customers.

Most of the CARS papers working in the representational view paradigm, consider only a few crucial contextual variables and focus on the development of novel CARS models and demonstrating their superior performance in using those variables. For example, \cite{HaririMobasherReviewMining2011} used only one contextual variable in hotels application, \cite{Tuzhilin2005} relied on only three contextual variables to recommend movies, \cite{AciarMiningCI,Li:2010:CRB:1944566.1944645} used only three variables to recommend restaurants. A popular CARS dataset LDOS-CoMoDa \cite{Odic2013PredictingAD} contains only twelve variables in movie recommendation application. An app recommendation dataset Frappe \cite{baltrunas2015frappe} provides only four contextual variables. All these prior works and methods designed for aforementioned datsets consider only a limited set of contextual variables ignoring a long list of contextual factors that could potentially affect user's preferences. 
In this paper, we study the long list of contextual variables forming the Long Tail of Context (LTC). First, we investigate if LTC exists in recommendation applications. Second, we study if the use of contextual variables from the LTC matters and helps to provide better recommendations.

\subsection{Conversational Recommender Systems}\label{sec:dialogue_systems}

Our work is also related to Conversational Recommender Systems (CRS) that are defined as software systems that support users in achieving recommendation-related goals through multi-turn dialogues \cite{CRS2021}. Most of the works in this area focus on understanding user intents (e.g., \cite{ZhaoAAAI2017}), modeling their profile (e.g., \cite{Thompson2004}), and tracking the state of the dialogue (e.g., \cite{RicciCRS2009}) to provide relevant recommendations. 
Different from these prior works, we deal with a novel task of extracting contextual information from user spoken inputs. We demonstrate that dialogue-based systems can be ``context-rich'' and contain a large variety of different types of contexts. We also show that extracted long tail of context helps to significantly improve recommendation performance.

\section{Extracting Long Tail of Context}
\label{sec:method}

In this section, we describe our dataset of dialogues between managers and clients, and the process of extracting the long tail of context from the transcribed texts of those dialogues.

\subsection{Dataset Description}\label{sec:dataset}
We work with a dataset $\mathcal{D}$ of transcribed dialogues between call-center managers of a large European Bank and its commercial clients. 
In these dialogues, managers offer clients $C_i\in \mathcal{C}$ various financial services $P_j\in \mathcal{P}$ provided by the company. 
Each dialogue $D_k\in\mathcal{D}$ contains up to three offers of services. After the call, managers also make a record of the outcome $r_{i,j}\in\{0,1\}$ indicating \textit{client's $C_i$ propensity to purchase} each of the discussed products $P_j$. Therefore, information about each of the dialogues $D_k$  contains its transcribed text $T(D_k)$, the list of offered products $\{P_a,P_b,P_c\}$, and the outcome for each of them $\{r_{i,a},r_{i,b},r_{i,c}\}$.
Our dialogues in $\mathcal{D}$ were recorded between July and September 2021.


We pre-processed dataset $\mathcal{D}$ to prepare it for our study. First, we removed very short dialogues where customers were not ready to talk and no products we offered. Second, we analyzed the data recorded by managers and filtered out the instances containing contradictory information, such as two offers of the same product during one dialogue with different recorded outcomes of client's propensity to purchase $r_{i,j}$. Second, we analyzed repeating calls to the same clients and combined them together. We filtered out those instances where the recorded outcome was different between the calls. As a result, we ended up with $247,365$ cleaned dialogues in the dataset.
Text $T(D_k)$ of each dialogue $D_k\in\mathcal{D}$ consist of customer's and manager's lines following each other. The average number of client's lines in our dialogues is $38.8$ with a median value of $15$. The average length of clients' lines is $11$ words. 


In our study, we focus on the four most important banking products that appear frequently in the dialogues, including \textit{Business Banking Account} with associated debit cards, \textit{Acquiring Service} when financial institution processes credit or debit card payments on behalf of a merchant,  \textit{Salary Service} to manage salaries and direct deposits for employees, and \textit{Leasing} service. 
Table~\ref{tab:datasetStats} reports the data statistics for these considered four products, including the number of dialogues (\#Dialogues) containing an offer of the product and percent of dialogues in which customers expressed their propensity to purchase that product (Purchase Propensity Rate).

\begin{table}[ht]
\resizebox{0.9\textwidth}{!}{%
   \begin{tabular}{|l|c|c|c|}
    \hline
    \textbf{Product Name} & \textbf{\#Dialogues} & \textbf{Purchase Propensity Rate} & \textbf{\% of dialogues with context} \\ \hline
    Business Banking Account &  $126,478$  & $24.28\%$ & $72.85\%$ \\\hline
    Acquiring Service     &  $58,750$   & $24.71\%$ & $75.87\%$ \\\hline
    Salary Service        &  $56,564$   & $41.13\%$ & $81.37\%$ \\\hline
    Leasing               &  $5,564$    & $18.44\%$ & $72.22\%$ \\\hline
\end{tabular}
}
\caption{Statistics of dialogues data for the considered bank products.} \label{tab:datasetStats}
\vspace{-1cm}
\end{table}

\subsection{Context Extraction Process}\label{sec:context_extraction}
To extract contextual information that was mentioned by the bank customers, we adapted the Context Parsing method described in \cite{bauman2021know} and applied it to our transcribed texts of dialogues. Context Parsing method was designed to parse contextual information from user reviews and it consists of five stages. It starts with a corpus of reviews, generates a set of syntactic phrases, filters, sorts, and analyzes them to identify the collection of contextual phrases, combines them into contextual variables, and, finally, marks all these contextual phases and variables in the specific reviews. 

In our study, we started with transcribed texts $T(D_k)$ of dialogues $D_k\in\mathcal{D}$ and considered only customers' lines as we are interested in contextual information mentioned by customers. We followed the same procedure as in \cite{bauman2021know} to generate the initial set of syntactic phrases up to four words long. Next, we filtered that set and selected only those phrases that appear in at least $50$ dialogues in $\mathcal{D}$. We also used t-test to identify and select only those phrases that significantly affect customer's propensity to purchase products when they are mentioned in dialogues. More specifically, for each phrase $w$ and product $P_j$, we compared purchase propensity rates calculated based on all dialogues offering $P_j$ and based on those dialogues offering $P_j$ that also mention $w$. As a result, we selected phrases that have such significant effect (with $p-value\ < \ 0.01$) on at least one of the considered products.  

Furthermore, we constructed embeddings for each of selected phrases using BERT model \cite{BERT2019} that was fine-tuned on corpora of banking-related texts using the standard masked language modelling approach. Our corpora included transcribed dialogues of bank managers offering various banking services to clients.
As a result, we obtained $768$-dimensional vector representations of generated phrases. We applied Principal Component Analysis (PCA) approach to compress that latent embedding space to $50$ dimensions. Next, we used clustering methods in the latent space to obtain groups of similar phrases. In our study, we tested two clustering methods, including Agglomerative Clustering \cite{AHC_Davidson_2005} and DBSCABN \cite{DBSCAN2017}, with different settings and compared the produced clusters based on the standard Silhouette Score measure \cite{rousseeuw1987silhouettes}. As a result, we ended up using DBSCAN method that produced $1,018$ clusters of phrases with the average size of $30.9$ phrases.

Next, following \cite{bauman2021know}, we calculated certain statistics for each of the clusters $t$, including the average customers' propensity rate (i.e., percent of dialogues containing phrases from $t$ where customer expressed their propensity to purchase), the average position of phrases from $t$ in dialogues, the average length of sentences containing phrases from $t$, percent of appearances in sentences in past tense, and percent of appearances with an attached sentiment. We used these statistics to further reduce the number of selected clusters. Finally, we engaged three domain experts among bank employees to do the fine-tuned selection of contextual clusters of phrases that refer to important contextual information in our application. Each of the experts evaluated the complete set of clusters and we used the majority of votes to select the final set of $216$ contextual clusters representing contextual variables. 
Each of the selected clusters contains phrases that reveal important contextual information mentioned by customers that has significant effect on customer's propensity to purchase for at least one of the considered products, such as ``customer's company is growing and they are about to open a new store,'' ``customer's company is about to hire new employees,'' ``customer's debit card is about to expire,'' ``they have an account in a different bank,'' ``they already have an acquiring terminal from another financial institution,'' ``customer's company is going through a court trial,'' ``customer is looking to buy a car for their company,'' ``type of customer's region, e.g. city vs small town or village,'' etc. Different from many other traditional CARS applications dealing with only a few crucial contextual variables, we work with a ``context-rich'' application having over two hundred contextual variables, the majority of which are infrequent and form the long tail of context.



Once we identified the list of contextual variables, we parse dialogues $\mathcal{D}$ and mark values of those variables in them. In particular, we set the value of the variable to $1$ if the dialogue contains the phrase from the corresponding contextual cluster. We also consider negations and for some contextual variables create a negative version, e.g., ``customer's company is NOT planning to hire new employees.'' The last column of Table~\ref{tab:datasetStats} reports the percentage of dialogues where at least one of our contextual variables were mentioned. 
As a result, for each dialogue $D_k\in\mathcal{D}$, we generated a vector $Context_{k}$ of contextual values that will be used for providing recommendations, as described in Section \ref{sec:analysis_settings}.

\section{Analysis and Experimental Settings}
\label{sec:analysis_settings}

To validate the power of the LTC, we conducted an empirical study on the dialogue-based system described in Section~\ref{sec:dataset}, where we provided context-aware recommendations by going deeper into the Long Tail of Context and thus using progressively more and more contextual variables extracted from the dialogues $\mathcal{D}$ and examining their effect on the quality of provided recommendations. 
In Section \ref{sec:cars_models}, we describe the method used for providing context-aware recommendations and in Section \ref{sec:eval_procedure}, we present the evaluation procedures.

\subsection{Context-Aware Recommendation Model}\label{sec:cars_models}


The recommender system used in our application recommends various financial services products to customers based on the customer data and on the previous historical interactions of the customer service representatives with them. Although the bank has a couple dozens of different products offered to commercial customers, in our study we focused on the following four products: (a) Business Banking Account; (b) Acquiring Service; (c) Salary Service; (d) Leasing, because they are of importance to the bank and its commercial customers, and appear frequently in the dialogues.
When recommending these products to the banking customers, we used the following information about them:
(a) type of the customer's business, its size and its various business activities, and
(b) transaction history of the customer, such as payments to the partners and customers of a business. 
This customer data was transformed into the latent space using Deep Learning methods and customer embeddings $Embed_i$ were generated using Contrastive Learning for Event Sequences with Self-Supervision (CoLES) approach \cite{CoLES:2022}. More specifically, CoLES splits customer's transaction history into short sequences, feeds them to the encoder, and applies contrastive learning loss to put sequences from the same customer next to each other in the embedding space while pushing sequences from different customers further away.

In addition to these customer embeddings, we also use the contextual information parsed from the dialogue systems as described in Section \ref{sec:context_extraction}, in order to recommend banking products to the customers as follows.
For each customer $C_i$ and each product $P_j$, we predict customer's propensity to purchase that product $r_{i,j,k}$ if that would be recommended within the context $Context_{k}$ associated with the dialogue $D_k$ (this association is parsed from $T(D_k)$ as described in Section \ref{sec:context_extraction}).
After producing such predictions $\hat{r}_{i,j,k}$ for each customer and product in a given context, we determine the most relevant products for each customer and the call center manager suggests those products to them.

To predict that propensity to purchase $r_{i,j,k}$, we train a classification model $\mathcal{M}$ for each considered product as follows. We start by selecting dialogues $\mathcal{D}^j$ where product $P_j$ was offered. Each of the dialogues $D_k\in\mathcal{D}^j$ is represented by corresponding customer embeddings $Embed_i$ and vector of contextual values $Context_{k}$ extracted from its text. We use these embeddings and contextual vectors as features to train model $\mathcal{M}$. 
In our study, we utilize various types of classification models $\mathcal{M}$ to predict customers propensities to purchase, including
(1) \textit{Logistic Regression} \cite{bishop2006pattern} linear model that works well in many practical cases;
(2) \textit{Random Forest} \cite{bishop2006pattern} ensemble method that fits a number of decision tree classifiers;
(3) \textit{LightGBM} \cite{ke2017lightgbm} gradient boosting framework that uses tree based learning algorithms;
(4) \textit{Factorization Machine} \cite{rendle2010factorization} model that is designed to capture interactions between features within high dimensional sparse datasets; and 
(5) \textit{AutoML} \cite{AutoML2021} that was specifically designed to work with ML problems of large financial services companies.
Once model $\mathcal{M}$ is trained, we predict propensity to purchase $r_{i,j,k}$ of customer $C_i$ and product $P_j$ within the current $Context_{k}$ by feeding the customer embedding $Embed_i$ and the contextual information $Context_{k}$ into model $\mathcal{M}$.

Note that the recommendation modeling approach we use in this paper relies on classification, that is similar to the ones described in \cite{amatriain2011data, musto2022semantics} and differs from the classical collaborative filtering methods \cite{Koren2022}. In our settings, the bank is interested to identify potential clients for each of the products and our approach allows to analyze and add/remove products independently.

\subsection{Evaluation Procedure for the Long Tail of Context}\label{sec:eval_procedure}





\begin{figure}[ht]
  \includegraphics[width=0.49\linewidth]{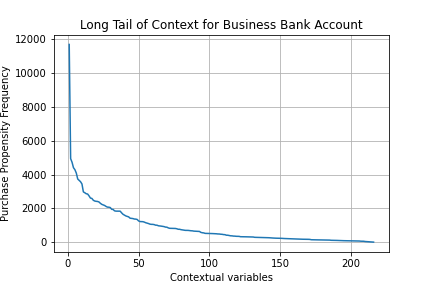}
  \includegraphics[width=0.49\linewidth]{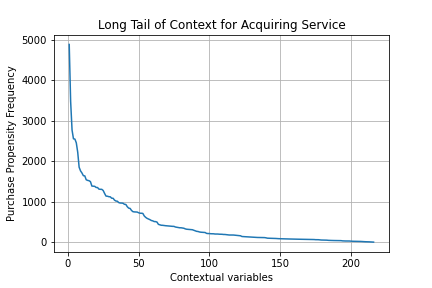}
\caption{Long Tail of Context based on its Purchase Propensity Frequency for Business Bank Account (left) and Aquaring Service (right).} \label{fig:long_tail_histogram}
\end{figure}

We evaluate how much the Long Tail of Context contributes to the quality of provided recommendations as follows.
First, we sort all the contextual variables based on their ``importance'' and then from the head of this distribution select progressively larger set of contextual variables that we use during the recommendation process. 
We sort the contextual variables in our application according to the following two criteria:
(1) the \textit{Purchase Propensity Frequency}, i.e., the number of times customers expresses their propensity to purchase product ($r_{i,j,k}=1$) when particular context was mentioned in a dialogue; and 
(2) the \textit{Purchase Propensity Rate}, i.e., the percentage of the dialogues containing the contextual variable in which the customer accepted the offer. 
The distributions of the Purchase Propensity Frequency values for two considered products are displayed in Figure~\ref{fig:long_tail_histogram}. These frequencies decrease following the power law and, thus, the long list of infrequent variables form the Long Tail of Context (LTC). 

After sorting contextual variables using the aforementioned two criteria, we apply the quantile approach and progressively select the top $q$ contextual variables from this distribution ($q = \{10\%, 20\%, \dots, 100\%\}$).
We train recommendation models $\mathcal{M}$ (Section~\ref{sec:cars_models}) using customer embeddings and these selected $q$ contextual variables and use the standard F1 and ROC-AUC metrics to evaluate the performance of our recommendation model. 
We apply the 10-fold cross-validation approach to train and test the performance of our models based on dataset $\mathcal{D}$.

The purpose of the described quantile-based approach is to demonstrate the power of the Long Tail of contextual variables, i.e., that addition of the next batch of contextual variables (and thus going ``deeper'' into the Long Tail) significantly improves recommendations due to a larger set of contextual variables used in the model. 
As baselines in our study, we consider models trained without context (i.e., $q=0\%$) as well as with different heads of the context distribution, i.e., $q$ equal to 10\% and 20\%, and compare by how much the recommendation performance changes when we consider the Long Tail of Context with larger values of $q$.

\section{Results}
\label{sec:results}

\begin{figure}[ht]
  \includegraphics[width=0.49\linewidth]{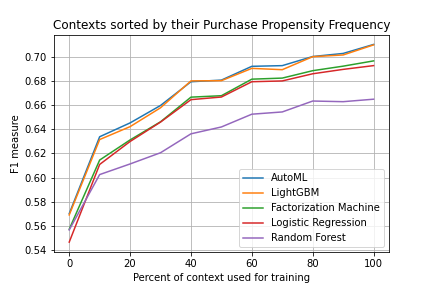}
  \includegraphics[width=0.49\linewidth]{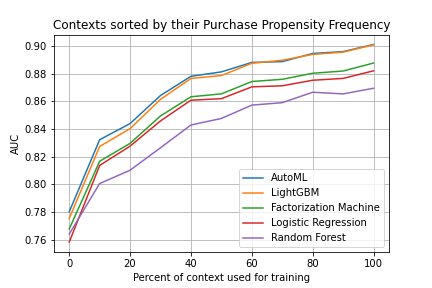}
\caption{Recommendation performance of models trained using different percent of contexts from the head of \textit{Purchase Propensity Frequency} distribution for \textit{Business Banking Account} service.} \label{fig:business_account}
\end{figure}
\begin{figure}[ht]
  \includegraphics[width=0.49\linewidth]{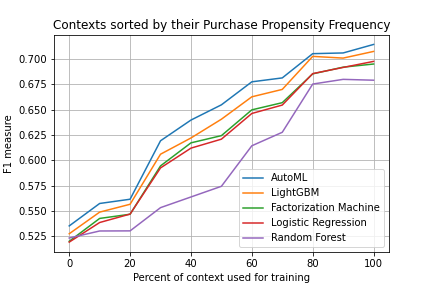}
  \includegraphics[width=0.49\linewidth]{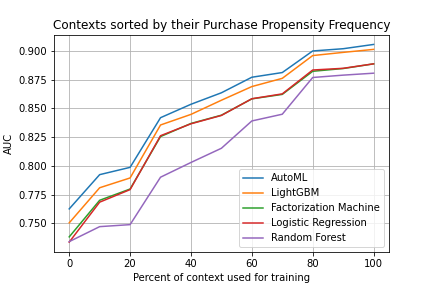}

\caption{Recommendation performance of models trained using different percent of contexts from the head of \textit{Purchase Propensity Frequency}  distribution for \textit{Acquiring} Service.} \label{fig:aquaring}
\end{figure}

The results of our study are presented in Figures \ref{fig:business_account} - \ref{fig:leasing}, where each figure shows how the recommendation performance metric (i.e., F1 measure and ROC AUC) changes when we consider progressively more contextual variables from the Long Tail in the model training process (ranging from no context at all to all the 100\% of all the contextual variables on the right of the x-axis) across different experimental settings. Figure~\ref{fig:business_account} focuses on the frequency-based sorting of contexts for the Business Banking Account and Figure~\ref{fig:aquaring} on the same sorting for the Acquiring Services.
Similar figures for other types of banking products are presented in Appendix A. 

As Figures \ref{fig:business_account} and \ref{fig:aquaring} demonstrate, the recommendation performance metrics tend to grow with the numbers of considered contextual variables incorporated into the recommendation models (from 0\% (no context) to 100\%). Furthermore, these figures show that the Head of the contextual variables distribution contributes very significantly to the performance improvements. This observation is in line with the numerous prior studies demonstrating that adding a few most important contextual variables leads to better recommendation performance \cite{Tuzhilin2005, HaririMobasherReviewMining2011}. Moreover, Figures \ref{fig:business_account} and \ref{fig:aquaring} also show that the Long Tail of Context also matters since significant performance improvements are observed when we move deeper into the Long Tail. For example, when we add only 10\% of all the contextual variables (20 variables) to the AutoML recommendation model, Figure \ref{fig:aquaring} shows 4.1\% of improvement of F1 measure vs. the case when no contextual variables are incorporated into the AutoML model; whereas when we add all 100\% of all the contextual variables (216 variables) into the AutoML model, we observe 33\% of performance improvement for the F1 measure for the Acquiring Service. 
This is an interesting and important new observation: many contextual variables in the Long Tail are idiosyncratic and often very specific that may also occur quite infrequently in the application, such as contextual variable ``customer's company is going through a court trial.''  However, once occurring in an application (e.g., in a dialogue with the customer), they play crucial role in the recommendation process. 
Furthermore, each individual variable may not be that important when considered separately. But there are many of such variables in the Long Tail of Context, and when considered \emph{together} they collectively lead to significant performance improvements, as Figures \ref{fig:business_account} and \ref{fig:aquaring} clearly demonstrate.

\begin{table}[ht]
\resizebox{0.95\textwidth}{!}{%
\begin{tabular}{|l|l|l|l|l|l|l|}
\hline
\multirow{2}{*}{\textbf{Product Name}}& \multirow{2}{*}{\textbf{Measure}} & \multicolumn{5}{c|}{\textbf{Percent of Context Used for Training}} \\ \cline{3-7}

& & No Context & 10\%  & 20\% & 50\% & 100\% \\ \hline
\multirow{2}{*}{Business Banking Account}& F1 & 0.57 & 0.634 (+11.1\%) & 0.645 (+13.2\%)  & 0.681 (+19.4\%)  & 0.71 (+24.5\%)\\
& AUC & 0.78 & 0.832 (+6.68\%) & 0.844 (+8.18\%)  & 0.881 (+13.0\%)  & 0.901 (+15.5\%)\\\hline
\multirow{2}{*}{Acquiring Service} & F1 & 0.535 & 0.557 (+4.14\%) & 0.562 (+4.92\%)  & 0.655 (+22.3\%)  & 0.714 (+33.5\%)\\
& AUC & 0.763 & 0.792 (+3.9\%) & 0.799 (+4.74\%)  & 0.863 (+13.2\%)  & 0.905 (+18.7\%)\\\hline
\multirow{2}{*}{Salary Service} & F1 & 0.65 & 0.662 (+1.8\%) & 0.672 (+3.33\%)  & 0.71 (+9.29\%)  & 0.744 (+14.5\%)\\
& AUC & 0.711 & 0.744 (+4.68\%) & 0.757 (+6.41\%)  & 0.826 (+16.2\%)  & 0.853 (+20.0\%)\\\hline
\multirow{2}{*}{Leasing} & F1 & 0.483 & 0.494 (+2.31\%) & 0.493 (+2.1\%)  & 0.529 (+9.56\%)  & 0.571 (+18.2\%)\\
& AUC & 0.757 & 0.762 (+0.697\%) & 0.764 (+0.952\%)  & 0.815 (+7.66\%)  & 0.851 (+12.4\%)\\\hline
\end{tabular}%
}
\caption{Prediction performance of AutoML model in terms of F1 and AUC measures trained using different percent of contexts from the head of distribution by their Purchase Propensity Frequency.} \label{tab:performance_improvements}
\vspace{-0.5cm}
\end{table}

Note that AutoML appears among the best performing models across various experimental settings (Figures \ref{fig:business_account}-\ref{fig:leasing}).
Therefore, for illustration purposes, Table~\ref{tab:performance_improvements} reports the numerical results for AutoML (based on Figures \ref{fig:business_account}-\ref{fig:leasing}) showing how much contextual variables in the Log Tail contribute to the recommendation performance improvements (in terms of the F1 and AUC measures) across four considered banking products. In particular, all the contextual variables including LTC (100\% of them) improve performance by 18.7\% for the AUC measure vis-\'a-vis 3.9\% for the contextual variables in the Head of the distribution (10\% of them) for the Acquiring Service. Similarly, for the Salary Service, using 10\% of contextual variables gives only  4.68\% improvement in terms of AUC measure, whereas using 100\% of LTC leads to 20\% recommendation performance improvement. 
Similar significant performance improvements are observed for other experimental settings, including different types of Long Tail distributions (Purchase Propensity Frequency and Rate), and recommendation models (non-AutoML), as Figures~\ref{fig:business_account}-\ref{fig:leasing} demonstrate. 
This observation demonstrates the strong power of the Long Tail of Context (LTC) in some recommender systems applications.

\section{Conclusion}
\label{sec:conclusion}
In this paper, we study those recommender systems applications having a large number of different types of contexts, such as dialogue-based systems with hundreds of different types of contextual variables. We demonstrate that these variables form a Long Tail of Context and that it matters in the sense that using all the contextual variables from the tail significantly improves performance of the CARS-based systems vis-\'a-vis using only a few most important contextual variables from the head of the distribution. We demonstrate this by studying
the application recommending various banking products to commercial customers within the context of dialogues initiated by customer service representatives of the bank, where we have identified 216 different types of contextual variables. 
Furthermore, we showed that the performance improvements of using all the 216 contextual variables range between 8\% and 29\% vs. the case of using the top-10 \% of most important variables across different experimental settings. 
These results are consistent across four considered banking products, five different classification models, and two types of sorting contextual variables. 
This demonstrates the importance of leveraging full contextual information in an application, as opposed to focusing only on a few most important types of context.

\begin{acks}
The authors would like to thank Kseniya Cheloshkina for the help with the data analysis and Alexey Grishanov for the help with running the experiments.
\end{acks}

\bibliographystyle{ACM-Reference-Format}
\bibliography{context_longtail}


\begin{thebibliography}{37}


\ifx \showCODEN    \undefined \def \showCODEN     #1{\unskip}     \fi
\ifx \showDOI      \undefined \def \showDOI       #1{#1}\fi
\ifx \showISBNx    \undefined \def \showISBNx     #1{\unskip}     \fi
\ifx \showISBNxiii \undefined \def \showISBNxiii  #1{\unskip}     \fi
\ifx \showISSN     \undefined \def \showISSN      #1{\unskip}     \fi
\ifx \showLCCN     \undefined \def \showLCCN      #1{\unskip}     \fi
\ifx \shownote     \undefined \def \shownote      #1{#1}          \fi
\ifx \showarticletitle \undefined \def \showarticletitle #1{#1}   \fi
\ifx \showURL      \undefined \def \showURL       {\relax}        \fi
\providecommand\bibfield[2]{#2}
\providecommand\bibinfo[2]{#2}
\providecommand\natexlab[1]{#1}
\providecommand\showeprint[2][]{arXiv:#2}

\bibitem[\protect\citeauthoryear{Aciar}{Aciar}{2010}]%
        {AciarMiningCI}
\bibfield{author}{\bibinfo{person}{Silvana Aciar}.}
  \bibinfo{year}{2010}\natexlab{}.
\newblock \showarticletitle{Mining Context Information from Consumer ’ s
  Reviews}. In \bibinfo{booktitle}{\emph{Workshop on Context-Aware Recommender
  System}}. \bibinfo{publisher}{ACM}.
\newblock


\bibitem[\protect\citeauthoryear{Adomavicius, Bauman, Tuzhilin, and
  Unger}{Adomavicius et~al\mbox{.}}{2022}]%
        {CARSChapter2022}
\bibfield{author}{\bibinfo{person}{Gediminas Adomavicius},
  \bibinfo{person}{Konstantin Bauman}, \bibinfo{person}{Alexander Tuzhilin},
  {and} \bibinfo{person}{Moshe Unger}.} \bibinfo{year}{2022}\natexlab{}.
\newblock \bibinfo{booktitle}{\emph{Context-Aware Recommender Systems: From
  Foundations to Recent DevelopmentsContext-aware recommender systems}}.
\newblock \bibinfo{publisher}{Springer US}, \bibinfo{address}{New York, NY},
  \bibinfo{pages}{211--250}.
\newblock
\showISBNx{978-1-0716-2197-4}
\urldef\tempurl%
\url{https://doi.org/10.1007/978-1-0716-2197-4_6}
\showDOI{\tempurl}


\bibitem[\protect\citeauthoryear{Adomavicius, Sankaranarayanan, Sen, and
  Tuzhilin}{Adomavicius et~al\mbox{.}}{2005}]%
        {Tuzhilin2005}
\bibfield{author}{\bibinfo{person}{Gediminas Adomavicius},
  \bibinfo{person}{Ramesh Sankaranarayanan}, \bibinfo{person}{Shahana Sen},
  {and} \bibinfo{person}{Alexander Tuzhilin}.} \bibinfo{year}{2005}\natexlab{}.
\newblock \showarticletitle{Incorporating Contextual Information in Recommender
  Systems Using a Multidimensional Approach}.
\newblock \bibinfo{journal}{\emph{ACM Trans. Inf. Syst.}} \bibinfo{volume}{23},
  \bibinfo{number}{1} (\bibinfo{date}{jan} \bibinfo{year}{2005}),
  \bibinfo{pages}{103–145}.
\newblock
\showISSN{1046-8188}
\urldef\tempurl%
\url{https://doi.org/10.1145/1055709.1055714}
\showDOI{\tempurl}


\bibitem[\protect\citeauthoryear{Amatriain, Oliver, Pujol,
  et~al\mbox{.}}{Amatriain et~al\mbox{.}}{2011}]%
        {amatriain2011data}
\bibfield{author}{\bibinfo{person}{Xavier Amatriain}, \bibinfo{person}{Nuria
  Oliver}, \bibinfo{person}{Josep~M Pujol}, {et~al\mbox{.}}}
  \bibinfo{year}{2011}\natexlab{}.
\newblock \showarticletitle{Data mining methods for recommender systems}.
\newblock In \bibinfo{booktitle}{\emph{Recommender systems handbook}}.
  \bibinfo{publisher}{Springer}, \bibinfo{pages}{39--71}.
\newblock


\bibitem[\protect\citeauthoryear{Babaev, Ovsov, Kireev, Ivanova, Gusev,
  Nazarov, and Tuzhilin}{Babaev et~al\mbox{.}}{2022}]%
        {CoLES:2022}
\bibfield{author}{\bibinfo{person}{Dmitrii Babaev}, \bibinfo{person}{Nikita
  Ovsov}, \bibinfo{person}{Ivan Kireev}, \bibinfo{person}{Maria Ivanova},
  \bibinfo{person}{Gleb Gusev}, \bibinfo{person}{Ivan Nazarov}, {and}
  \bibinfo{person}{Alexander Tuzhilin}.} \bibinfo{year}{2022}\natexlab{}.
\newblock \showarticletitle{CoLES: Contrastive Learning for Event Sequences
  with Self-Supervision}. In \bibinfo{booktitle}{\emph{Proceedings of the 2022
  International Conference on Management of Data}} (Philadelphia, PA, USA)
  \emph{(\bibinfo{series}{SIGMOD '22})}. \bibinfo{publisher}{Association for
  Computing Machinery}, \bibinfo{address}{New York, NY, USA},
  \bibinfo{pages}{1190–1199}.
\newblock
\showISBNx{9781450392495}
\urldef\tempurl%
\url{https://doi.org/10.1145/3514221.3526129}
\showDOI{\tempurl}


\bibitem[\protect\citeauthoryear{Baltrunas, Church, Karatzoglou, and
  Oliver}{Baltrunas et~al\mbox{.}}{2015}]%
        {baltrunas2015frappe}
\bibfield{author}{\bibinfo{person}{Linas Baltrunas}, \bibinfo{person}{Karen
  Church}, \bibinfo{person}{Alexandros Karatzoglou}, {and}
  \bibinfo{person}{Nuria Oliver}.} \bibinfo{year}{2015}\natexlab{}.
\newblock \showarticletitle{Frappe: Understanding the Usage and Perception of
  Mobile App Recommendations In-The-Wild}.
\newblock  (\bibinfo{year}{2015}).
\newblock
\showeprint[arxiv]{1505.03014}~[cs.IR]


\bibitem[\protect\citeauthoryear{Bauman and Tuzhilin}{Bauman and
  Tuzhilin}{2022}]%
        {bauman2021know}
\bibfield{author}{\bibinfo{person}{Konstantin Bauman} {and}
  \bibinfo{person}{Alexander Tuzhilin}.} \bibinfo{year}{2022}\natexlab{}.
\newblock \showarticletitle{Know Thy Context: Parsing Contextual Information
  from User Reviews for Recommendation Purposes}.
\newblock \bibinfo{journal}{\emph{Information Systems Research}}
  \bibinfo{volume}{33}, \bibinfo{number}{1} (\bibinfo{year}{2022}),
  \bibinfo{pages}{179--202}.
\newblock


\bibitem[\protect\citeauthoryear{Bishop and Nasrabadi}{Bishop and
  Nasrabadi}{2006}]%
        {bishop2006pattern}
\bibfield{author}{\bibinfo{person}{Christopher~M Bishop} {and}
  \bibinfo{person}{Nasser~M Nasrabadi}.} \bibinfo{year}{2006}\natexlab{}.
\newblock \bibinfo{booktitle}{\emph{Pattern recognition and machine learning}}.
  Vol.~\bibinfo{volume}{4}.
\newblock \bibinfo{publisher}{Springer}.
\newblock


\bibitem[\protect\citeauthoryear{Chen and Chen}{Chen and Chen}{2014}]%
        {10.1007/978-3-319-08786-3_6}
\bibfield{author}{\bibinfo{person}{Guanliang Chen} {and} \bibinfo{person}{Li
  Chen}.} \bibinfo{year}{2014}\natexlab{}.
\newblock \showarticletitle{Recommendation Based on Contextual Opinions}. In
  \bibinfo{booktitle}{\emph{User Modeling, Adaptation, and Personalization}},
  \bibfield{editor}{\bibinfo{person}{Vania Dimitrova}, \bibinfo{person}{Tsvi
  Kuflik}, \bibinfo{person}{David Chin}, \bibinfo{person}{Francesco Ricci},
  \bibinfo{person}{Peter Dolog}, {and} \bibinfo{person}{Geert-Jan Houben}}
  (Eds.). \bibinfo{publisher}{Springer International Publishing},
  \bibinfo{address}{Cham}, \bibinfo{pages}{61--73}.
\newblock


\bibitem[\protect\citeauthoryear{Chen and Chen}{Chen and Chen}{2015}]%
        {Chen:2015}
\bibfield{author}{\bibinfo{person}{Guanliang Chen} {and} \bibinfo{person}{Li
  Chen}.} \bibinfo{year}{2015}\natexlab{}.
\newblock \showarticletitle{Augmenting Service Recommender Systems by
  Incorporating Contextual Opinions from User Reviews}.
\newblock \bibinfo{journal}{\emph{User Modeling and User-Adapted Interaction}}
  \bibinfo{volume}{25}, \bibinfo{number}{3} (\bibinfo{date}{Aug.}
  \bibinfo{year}{2015}), \bibinfo{pages}{295--329}.
\newblock


\bibitem[\protect\citeauthoryear{Davidson and Ravi}{Davidson and Ravi}{2005}]%
        {AHC_Davidson_2005}
\bibfield{author}{\bibinfo{person}{Ian Davidson} {and} \bibinfo{person}{S.~S.
  Ravi}.} \bibinfo{year}{2005}\natexlab{}.
\newblock \showarticletitle{Agglomerative Hierarchical Clustering with
  Constraints: Theoretical and Empirical Results}
  \emph{(\bibinfo{series}{ECMLPKDD'05})}. \bibinfo{publisher}{Springer-Verlag},
  \bibinfo{address}{Berlin, Heidelberg}, \bibinfo{pages}{59–70}.
\newblock
\showISBNx{3540292446}


\bibitem[\protect\citeauthoryear{del Carmen Rodr{\'\i}guez-Hern{\'a}ndez and
  Ilarri}{del Carmen Rodr{\'\i}guez-Hern{\'a}ndez and Ilarri}{2021}]%
        {del2021ai}
\bibfield{author}{\bibinfo{person}{Mar{\'\i}a del Carmen
  Rodr{\'\i}guez-Hern{\'a}ndez} {and} \bibinfo{person}{Sergio Ilarri}.}
  \bibinfo{year}{2021}\natexlab{}.
\newblock \showarticletitle{AI-based mobile context-aware recommender systems
  from an information management perspective: Progress and directions}.
\newblock \bibinfo{journal}{\emph{Knowledge-Based Systems}}
  \bibinfo{volume}{215} (\bibinfo{year}{2021}), \bibinfo{pages}{106740}.
\newblock


\bibitem[\protect\citeauthoryear{Devlin, Chang, Lee, and Toutanova}{Devlin
  et~al\mbox{.}}{2019}]%
        {BERT2019}
\bibfield{author}{\bibinfo{person}{Jacob Devlin}, \bibinfo{person}{Ming{-}Wei
  Chang}, \bibinfo{person}{Kenton Lee}, {and} \bibinfo{person}{Kristina
  Toutanova}.} \bibinfo{year}{2019}\natexlab{}.
\newblock \showarticletitle{{BERT:} Pre-training of Deep Bidirectional
  Transformers for Language Understanding}. In
  \bibinfo{booktitle}{\emph{Proceedings of the 2019 Conference of the North
  American Chapter of the Association for Computational Linguistics: Human
  Language Technologies, {NAACL-HLT} 2019, Minneapolis, MN, USA, June 2-7,
  2019, Volume 1 (Long and Short Papers)}},
  \bibfield{editor}{\bibinfo{person}{Jill Burstein}, \bibinfo{person}{Christy
  Doran}, {and} \bibinfo{person}{Thamar Solorio}} (Eds.).
  \bibinfo{publisher}{Association for Computational Linguistics},
  \bibinfo{pages}{4171--4186}.
\newblock
\urldef\tempurl%
\url{https://doi.org/10.18653/v1/n19-1423}
\showDOI{\tempurl}


\bibitem[\protect\citeauthoryear{Ding, Tang, Liu, Xu, Zhang, Shi, Jiang, and
  Shen}{Ding et~al\mbox{.}}{2019}]%
        {Ding2019}
\bibfield{author}{\bibinfo{person}{Xichen Ding}, \bibinfo{person}{Jie Tang},
  \bibinfo{person}{Tracy Liu}, \bibinfo{person}{Cheng Xu},
  \bibinfo{person}{Yaping Zhang}, \bibinfo{person}{Feng Shi},
  \bibinfo{person}{Qixia Jiang}, {and} \bibinfo{person}{Dan Shen}.}
  \bibinfo{year}{2019}\natexlab{}.
\newblock \showarticletitle{Infer Implicit Contexts in Real-Time
  Online-to-Offline Recommendation}. In \bibinfo{booktitle}{\emph{Proceedings
  of the 25th ACM SIGKDD International Conference on Knowledge Discovery \&
  Data Mining}} (Anchorage, AK, USA) \emph{(\bibinfo{series}{KDD ’19})}.
  \bibinfo{publisher}{Association for Computing Machinery},
  \bibinfo{address}{New York, NY, USA}, \bibinfo{pages}{2336–2346}.
\newblock
\showISBNx{9781450362016}
\urldef\tempurl%
\url{https://doi.org/10.1145/3292500.3330716}
\showDOI{\tempurl}


\bibitem[\protect\citeauthoryear{Dourish}{Dourish}{2004}]%
        {Dourish:2004}
\bibfield{author}{\bibinfo{person}{Paul Dourish}.}
  \bibinfo{year}{2004}\natexlab{}.
\newblock \showarticletitle{What We Talk About when We Talk About Context}.
\newblock \bibinfo{journal}{\emph{Personal Ubiquitous Comput.}}
  \bibinfo{volume}{8}, \bibinfo{number}{1} (\bibinfo{date}{Feb.}
  \bibinfo{year}{2004}), \bibinfo{pages}{19--30}.
\newblock


\bibitem[\protect\citeauthoryear{Dragone, Mehrotra, and Lalmas}{Dragone
  et~al\mbox{.}}{2019}]%
        {DragoneWWW2019}
\bibfield{author}{\bibinfo{person}{Paolo Dragone}, \bibinfo{person}{Rishabh
  Mehrotra}, {and} \bibinfo{person}{Mounia Lalmas}.}
  \bibinfo{year}{2019}\natexlab{}.
\newblock \showarticletitle{Deriving User- and Content-Specific Rewards for
  Contextual Bandits}. In \bibinfo{booktitle}{\emph{The World Wide Web
  Conference}} (San Francisco, CA, USA) \emph{(\bibinfo{series}{WWW ’19})}.
  \bibinfo{publisher}{Association for Computing Machinery},
  \bibinfo{address}{New York, NY, USA}, \bibinfo{pages}{2680–2686}.
\newblock


\bibitem[\protect\citeauthoryear{Hansen, Hansen, Maystre, Mehrotra, Brost,
  Tomasi, and Lalmas}{Hansen et~al\mbox{.}}{2020}]%
        {hansen2020contextual}
\bibfield{author}{\bibinfo{person}{Casper Hansen}, \bibinfo{person}{Christian
  Hansen}, \bibinfo{person}{Lucas Maystre}, \bibinfo{person}{Rishabh Mehrotra},
  \bibinfo{person}{Brian Brost}, \bibinfo{person}{Federico Tomasi}, {and}
  \bibinfo{person}{Mounia Lalmas}.} \bibinfo{year}{2020}\natexlab{}.
\newblock \showarticletitle{Contextual and sequential user embeddings for
  large-scale music recommendation}. In \bibinfo{booktitle}{\emph{Fourteenth
  ACM Conference on Recommender Systems}}. \bibinfo{pages}{53--62}.
\newblock


\bibitem[\protect\citeauthoryear{Hariri, Mobasher, Burke, and Zheng}{Hariri
  et~al\mbox{.}}{2011}]%
        {HaririMobasherReviewMining2011}
\bibfield{author}{\bibinfo{person}{Negar Hariri}, \bibinfo{person}{Bamshad
  Mobasher}, \bibinfo{person}{Robin Burke}, {and} \bibinfo{person}{Yong
  Zheng}.} \bibinfo{year}{2011}\natexlab{}.
\newblock \showarticletitle{Context-Aware Recommendation Based On Review
  Mining}. In \bibinfo{booktitle}{\emph{Proceedings of the 9th Workshop on
  Intelligent Techniques for Web Personalization {\&} Recommender Systems,
  ITWP@IJCAI 2011, Barcelona, Spain, July 16, 2011}}.
\newblock


\bibitem[\protect\citeauthoryear{Jannach, Manzoor, Cai, and Chen}{Jannach
  et~al\mbox{.}}{2021}]%
        {CRS2021}
\bibfield{author}{\bibinfo{person}{Dietmar Jannach}, \bibinfo{person}{Ahtsham
  Manzoor}, \bibinfo{person}{Wanling Cai}, {and} \bibinfo{person}{Li Chen}.}
  \bibinfo{year}{2021}\natexlab{}.
\newblock \showarticletitle{A Survey on Conversational Recommender Systems}.
\newblock \bibinfo{journal}{\emph{ACM Comput. Surv.}} \bibinfo{volume}{54},
  \bibinfo{number}{5}, Article \bibinfo{articleno}{105} (\bibinfo{date}{may}
  \bibinfo{year}{2021}), \bibinfo{numpages}{36}~pages.
\newblock
\showISSN{0360-0300}
\urldef\tempurl%
\url{https://doi.org/10.1145/3453154}
\showDOI{\tempurl}


\bibitem[\protect\citeauthoryear{Ke, Meng, Finley, Wang, Chen, Ma, Ye, and
  Liu}{Ke et~al\mbox{.}}{2017}]%
        {ke2017lightgbm}
\bibfield{author}{\bibinfo{person}{Guolin Ke}, \bibinfo{person}{Qi Meng},
  \bibinfo{person}{Thomas Finley}, \bibinfo{person}{Taifeng Wang},
  \bibinfo{person}{Wei Chen}, \bibinfo{person}{Weidong Ma},
  \bibinfo{person}{Qiwei Ye}, {and} \bibinfo{person}{Tie-Yan Liu}.}
  \bibinfo{year}{2017}\natexlab{}.
\newblock \showarticletitle{Lightgbm: A highly efficient gradient boosting
  decision tree}.
\newblock \bibinfo{journal}{\emph{Advances in neural information processing
  systems}}  \bibinfo{volume}{30} (\bibinfo{year}{2017}).
\newblock


\bibitem[\protect\citeauthoryear{Koren, Rendle, and Bell}{Koren
  et~al\mbox{.}}{2022}]%
        {Koren2022}
\bibfield{author}{\bibinfo{person}{Yehuda Koren}, \bibinfo{person}{Steffen
  Rendle}, {and} \bibinfo{person}{Robert Bell}.}
  \bibinfo{year}{2022}\natexlab{}.
\newblock \bibinfo{booktitle}{\emph{Advances in Collaborative Filtering}}.
\newblock \bibinfo{publisher}{Springer US}, \bibinfo{address}{New York, NY},
  \bibinfo{pages}{91--142}.
\newblock
\showISBNx{978-1-0716-2197-4}
\urldef\tempurl%
\url{https://doi.org/10.1007/978-1-0716-2197-4_3}
\showDOI{\tempurl}


\bibitem[\protect\citeauthoryear{Li, Nie, Zhang, Wang, Yan, and Weng}{Li
  et~al\mbox{.}}{2010}]%
        {Li:2010:CRB:1944566.1944645}
\bibfield{author}{\bibinfo{person}{Yize Li}, \bibinfo{person}{Jiazhong Nie},
  \bibinfo{person}{Yi Zhang}, \bibinfo{person}{Bingqing Wang},
  \bibinfo{person}{Baoshi Yan}, {and} \bibinfo{person}{Fuliang Weng}.}
  \bibinfo{year}{2010}\natexlab{}.
\newblock \showarticletitle{Contextual Recommendation Based on Text Mining}. In
  \bibinfo{booktitle}{\emph{Proceedings of the 23rd International Conference on
  Computational Linguistics: Posters}} (Beijing, China)
  \emph{(\bibinfo{series}{COLING '10})}. \bibinfo{publisher}{Association for
  Computational Linguistics}, \bibinfo{address}{Stroudsburg, PA, USA},
  \bibinfo{pages}{692--700}.
\newblock


\bibitem[\protect\citeauthoryear{Mahmood and Ricci}{Mahmood and Ricci}{2009}]%
        {RicciCRS2009}
\bibfield{author}{\bibinfo{person}{Tariq Mahmood} {and}
  \bibinfo{person}{Francesco Ricci}.} \bibinfo{year}{2009}\natexlab{}.
\newblock \showarticletitle{Improving Recommender Systems with Adaptive
  Conversational Strategies}. In \bibinfo{booktitle}{\emph{Proceedings of the
  20th ACM Conference on Hypertext and Hypermedia}} (Torino, Italy)
  \emph{(\bibinfo{series}{HT '09})}. \bibinfo{publisher}{Association for
  Computing Machinery}, \bibinfo{address}{New York, NY, USA},
  \bibinfo{pages}{73–82}.
\newblock
\showISBNx{9781605584867}
\urldef\tempurl%
\url{https://doi.org/10.1145/1557914.1557930}
\showDOI{\tempurl}


\bibitem[\protect\citeauthoryear{McInerney, Lacker, Hansen, Higley, Bouchard,
  Gruson, and Mehrotra}{McInerney et~al\mbox{.}}{2018}]%
        {McInerneyRecSys2019}
\bibfield{author}{\bibinfo{person}{James McInerney}, \bibinfo{person}{Benjamin
  Lacker}, \bibinfo{person}{Samantha Hansen}, \bibinfo{person}{Karl Higley},
  \bibinfo{person}{Hugues Bouchard}, \bibinfo{person}{Alois Gruson}, {and}
  \bibinfo{person}{Rishabh Mehrotra}.} \bibinfo{year}{2018}\natexlab{}.
\newblock \showarticletitle{Explore, Exploit, and Explain: Personalizing
  Explainable Recommendations with Bandits}. In
  \bibinfo{booktitle}{\emph{Proceedings of the 12th ACM Conference on
  Recommender Systems}} (Vancouver, British Columbia, Canada)
  \emph{(\bibinfo{series}{RecSys ’18})}. \bibinfo{publisher}{Association for
  Computing Machinery}, \bibinfo{address}{New York, NY, USA},
  \bibinfo{pages}{31–39}.
\newblock


\bibitem[\protect\citeauthoryear{Musto, Gemmis, Lops, Narducci, and
  Semeraro}{Musto et~al\mbox{.}}{2022}]%
        {musto2022semantics}
\bibfield{author}{\bibinfo{person}{Cataldo Musto}, \bibinfo{person}{Marco~de
  Gemmis}, \bibinfo{person}{Pasquale Lops}, \bibinfo{person}{Fedelucio
  Narducci}, {and} \bibinfo{person}{Giovanni Semeraro}.}
  \bibinfo{year}{2022}\natexlab{}.
\newblock \showarticletitle{Semantics and content-based recommendations}.
\newblock In \bibinfo{booktitle}{\emph{Recommender systems handbook}}.
  \bibinfo{publisher}{Springer}, \bibinfo{pages}{251--298}.
\newblock


\bibitem[\protect\citeauthoryear{Odic, Tkalcic, Tasic, and Kosir}{Odic
  et~al\mbox{.}}{2013}]%
        {Odic2013PredictingAD}
\bibfield{author}{\bibinfo{person}{Ante Odic}, \bibinfo{person}{Marko Tkalcic},
  \bibinfo{person}{Jurij~F. Tasic}, {and} \bibinfo{person}{Andrej Kosir}.}
  \bibinfo{year}{2013}\natexlab{}.
\newblock \showarticletitle{Predicting and Detecting the Relevant Contextual
  Information in a Movie-Recommender System}.
\newblock \bibinfo{journal}{\emph{Interacting with Computers}}
  \bibinfo{volume}{25} (\bibinfo{year}{2013}), \bibinfo{pages}{74--90}.
\newblock


\bibitem[\protect\citeauthoryear{Rendle}{Rendle}{2010}]%
        {rendle2010factorization}
\bibfield{author}{\bibinfo{person}{Steffen Rendle}.}
  \bibinfo{year}{2010}\natexlab{}.
\newblock \showarticletitle{Factorization machines}. In
  \bibinfo{booktitle}{\emph{2010 IEEE International conference on data
  mining}}. IEEE, \bibinfo{pages}{995--1000}.
\newblock


\bibitem[\protect\citeauthoryear{Rousseeuw}{Rousseeuw}{1987}]%
        {rousseeuw1987silhouettes}
\bibfield{author}{\bibinfo{person}{Peter~J Rousseeuw}.}
  \bibinfo{year}{1987}\natexlab{}.
\newblock \showarticletitle{Silhouettes: a graphical aid to the interpretation
  and validation of cluster analysis}.
\newblock \bibinfo{journal}{\emph{Journal of computational and applied
  mathematics}}  \bibinfo{volume}{20} (\bibinfo{year}{1987}),
  \bibinfo{pages}{53--65}.
\newblock


\bibitem[\protect\citeauthoryear{Schubert, Sander, Ester, Kriegel, and
  Xu}{Schubert et~al\mbox{.}}{2017}]%
        {DBSCAN2017}
\bibfield{author}{\bibinfo{person}{Erich Schubert}, \bibinfo{person}{J\"{o}rg
  Sander}, \bibinfo{person}{Martin Ester}, \bibinfo{person}{Hans~Peter
  Kriegel}, {and} \bibinfo{person}{Xiaowei Xu}.}
  \bibinfo{year}{2017}\natexlab{}.
\newblock \showarticletitle{DBSCAN Revisited, Revisited: Why and How You Should
  (Still) Use DBSCAN}.
\newblock \bibinfo{journal}{\emph{ACM Trans. Database Syst.}}
  \bibinfo{volume}{42}, \bibinfo{number}{3}, Article \bibinfo{articleno}{19}
  (\bibinfo{date}{jul} \bibinfo{year}{2017}), \bibinfo{numpages}{21}~pages.
\newblock
\showISSN{0362-5915}
\urldef\tempurl%
\url{https://doi.org/10.1145/3068335}
\showDOI{\tempurl}


\bibitem[\protect\citeauthoryear{Selvi and Sivasankar}{Selvi and
  Sivasankar}{2019}]%
        {DSBDA2018}
\bibfield{author}{\bibinfo{person}{C. Selvi} {and} \bibinfo{person}{E.
  Sivasankar}.} \bibinfo{year}{2019}\natexlab{}.
\newblock \showarticletitle{An Efficient Context-Aware Music Recommendation
  Based on Emotion and Time Context}. In \bibinfo{booktitle}{\emph{Data Science
  and Big Data Analytics}}, \bibfield{editor}{\bibinfo{person}{Durgesh~Kumar
  Mishra}, \bibinfo{person}{Xin-She Yang}, {and} \bibinfo{person}{Aynur Unal}}
  (Eds.). \bibinfo{pages}{215--228}.
\newblock


\bibitem[\protect\citeauthoryear{Thompson, G\"{o}ker, and Langley}{Thompson
  et~al\mbox{.}}{2004}]%
        {Thompson2004}
\bibfield{author}{\bibinfo{person}{Cynthia~A. Thompson},
  \bibinfo{person}{Mehmet~H. G\"{o}ker}, {and} \bibinfo{person}{Pat Langley}.}
  \bibinfo{year}{2004}\natexlab{}.
\newblock \showarticletitle{A Personalized System for Conversational
  Recommendations}.
\newblock \bibinfo{journal}{\emph{J. Artif. Int. Res.}} \bibinfo{volume}{21},
  \bibinfo{number}{1} (\bibinfo{date}{mar} \bibinfo{year}{2004}),
  \bibinfo{pages}{393–428}.
\newblock
\showISSN{1076-9757}


\bibitem[\protect\citeauthoryear{Unger, Tuzhilin, and Livne}{Unger
  et~al\mbox{.}}{2020}]%
        {unger2020context}
\bibfield{author}{\bibinfo{person}{Moshe Unger}, \bibinfo{person}{Alexander
  Tuzhilin}, {and} \bibinfo{person}{Amit Livne}.}
  \bibinfo{year}{2020}\natexlab{}.
\newblock \showarticletitle{Context-Aware Recommendations Based on Deep
  Learning Frameworks}.
\newblock \bibinfo{journal}{\emph{ACM Transactions on Management Information
  Systems (TMIS)}} \bibinfo{volume}{11}, \bibinfo{number}{2}
  (\bibinfo{year}{2020}), \bibinfo{pages}{1--15}.
\newblock


\bibitem[\protect\citeauthoryear{Vakhrushev, Ryzhkov, Savchenko, Simakov,
  Damdinov, and Tuzhilin}{Vakhrushev et~al\mbox{.}}{2021}]%
        {AutoML2021}
\bibfield{author}{\bibinfo{person}{Anton Vakhrushev},
  \bibinfo{person}{Alexander Ryzhkov}, \bibinfo{person}{Maxim Savchenko},
  \bibinfo{person}{Dmitry Simakov}, \bibinfo{person}{Rinchin Damdinov}, {and}
  \bibinfo{person}{Alexander Tuzhilin}.} \bibinfo{year}{2021}\natexlab{}.
\newblock \bibinfo{title}{LightAutoML: AutoML Solution for a Large Financial
  Services Ecosystem}.
\newblock
\newblock
\urldef\tempurl%
\url{https://doi.org/10.48550/ARXIV.2109.01528}
\showDOI{\tempurl}


\bibitem[\protect\citeauthoryear{Villegas, S{\'a}nchez, D{\'\i}az-Cely, and
  Tamura}{Villegas et~al\mbox{.}}{2018}]%
        {villegas2018characterizing}
\bibfield{author}{\bibinfo{person}{Norha~M Villegas}, \bibinfo{person}{Cristian
  S{\'a}nchez}, \bibinfo{person}{Javier D{\'\i}az-Cely}, {and}
  \bibinfo{person}{Gabriel Tamura}.} \bibinfo{year}{2018}\natexlab{}.
\newblock \showarticletitle{Characterizing context-aware recommender systems: A
  systematic literature review}.
\newblock \bibinfo{journal}{\emph{Knowledge-Based Systems}}
  \bibinfo{volume}{140} (\bibinfo{year}{2018}), \bibinfo{pages}{173--200}.
\newblock


\bibitem[\protect\citeauthoryear{Wu, Zhao, Zhang, Meng, Zhang, Zhang, and
  Sun}{Wu et~al\mbox{.}}{2017}]%
        {Wu:2017:IPT:3103645.3103886}
\bibfield{author}{\bibinfo{person}{Wenmin Wu}, \bibinfo{person}{Jianli Zhao},
  \bibinfo{person}{Chunsheng Zhang}, \bibinfo{person}{Fang Meng},
  \bibinfo{person}{Zeli Zhang}, \bibinfo{person}{Yang Zhang}, {and}
  \bibinfo{person}{Qiuxia Sun}.} \bibinfo{year}{2017}\natexlab{}.
\newblock \showarticletitle{Improving Performance of Tensor-based Context-aware
  Recommenders Using Bias Tensor Factorization with Context Feature
  Auto-encoding}.
\newblock \bibinfo{journal}{\emph{Know.-Based Syst.}} \bibinfo{volume}{128},
  \bibinfo{number}{C} (\bibinfo{date}{July} \bibinfo{year}{2017}),
  \bibinfo{pages}{71--77}.
\newblock


\bibitem[\protect\citeauthoryear{Xin, Chen, He, Wang, Ding, and Jose}{Xin
  et~al\mbox{.}}{2019}]%
        {xin2019cfm}
\bibfield{author}{\bibinfo{person}{Xin Xin}, \bibinfo{person}{Bo Chen},
  \bibinfo{person}{Xiangnan He}, \bibinfo{person}{Dong Wang},
  \bibinfo{person}{Yue Ding}, {and} \bibinfo{person}{Joemon Jose}.}
  \bibinfo{year}{2019}\natexlab{}.
\newblock \showarticletitle{CFM: convolutional factorization machines for
  context-aware recommendation}. In \bibinfo{booktitle}{\emph{Proceedings of
  the 28th International Joint Conference on Artificial Intelligence}}. AAAI
  Press, \bibinfo{pages}{3926--3932}.
\newblock


\bibitem[\protect\citeauthoryear{Yan, Duan, Chen, Zhou, Zhou, and Li}{Yan
  et~al\mbox{.}}{2017}]%
        {ZhaoAAAI2017}
\bibfield{author}{\bibinfo{person}{Zhao Yan}, \bibinfo{person}{Nan Duan},
  \bibinfo{person}{Peng Chen}, \bibinfo{person}{Ming Zhou},
  \bibinfo{person}{Jianshe Zhou}, {and} \bibinfo{person}{Zhoujun Li}.}
  \bibinfo{year}{2017}\natexlab{}.
\newblock \showarticletitle{Building Task-Oriented Dialogue Systems for Online
  Shopping}. In \bibinfo{booktitle}{\emph{Proceedings of the Thirty-First AAAI
  Conference on Artificial Intelligence}} (San Francisco, California, USA)
  \emph{(\bibinfo{series}{AAAI'17})}. \bibinfo{publisher}{AAAI Press},
  \bibinfo{pages}{4618–4625}.
\newblock


\end{thebibliography}

\newpage
\appendix

\section{Additional Plots reporting Results}\label{sec:appendix}
\begin{figure}[ht]
  \includegraphics[width=0.49\linewidth]{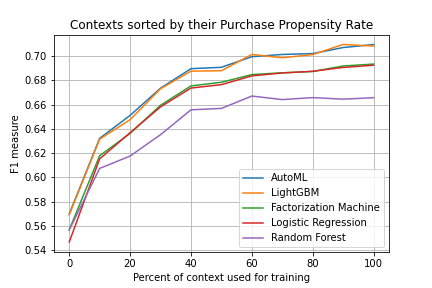}
  \includegraphics[width=0.49\linewidth]{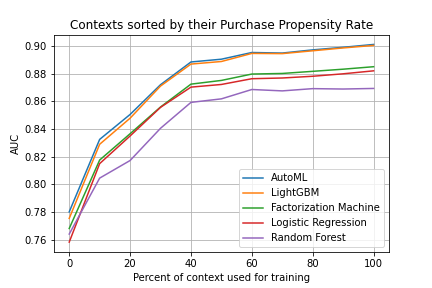}
\caption{Recommendation performance of models trained using different percent of contexts from the head of \textit{Purchase Propensity Rate} distribution for \textit{Business Banking Account} service.} \label{fig:business_account_2}
\end{figure}

\begin{figure}[ht]
  \includegraphics[width=0.49\linewidth]{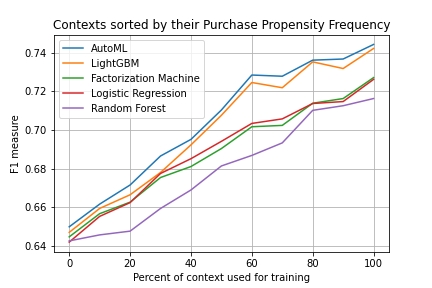}
  \includegraphics[width=0.49\linewidth]{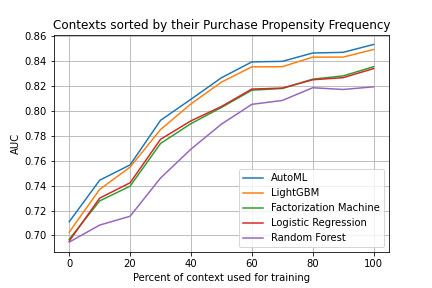}
  \includegraphics[width=0.49\linewidth]{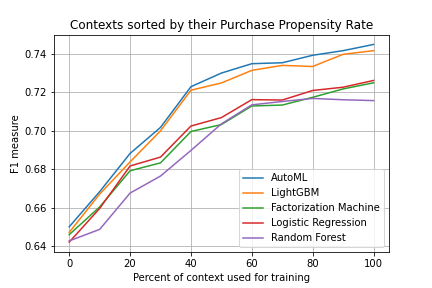}
  \includegraphics[width=0.49\linewidth]{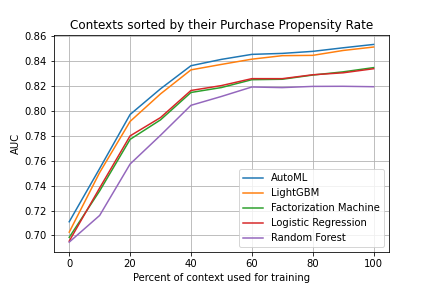}
\caption{Recommendation performance of models trained using different percent of contexts from the head of \textit{Purchase Propensity Frequency} and \textit{Purchase Propensity Rate} distributions for \textit{Salaries} Service.} \label{fig:salaries}
\end{figure}

\begin{figure}[ht]
  \includegraphics[width=0.49\linewidth]{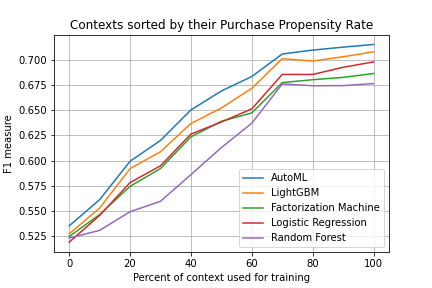}
  \includegraphics[width=0.49\linewidth]{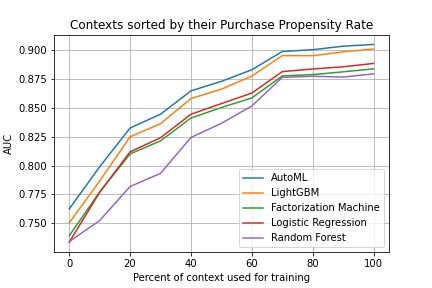}

\caption{Recommendation performance of models trained using different percent of contexts from the head of \textit{Purchase Propensity Rate} distribution for Acquiring Service.} \label{fig:aquaring_2}
\end{figure}

\begin{figure}[ht]
  \includegraphics[width=0.49\linewidth]{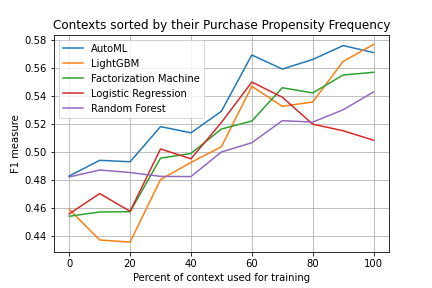}
  \includegraphics[width=0.49\linewidth]{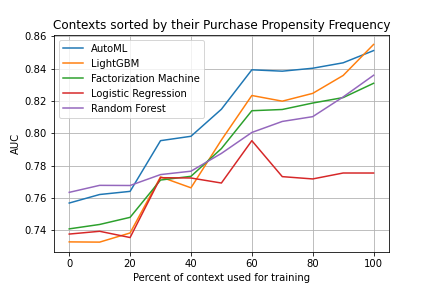}
  \includegraphics[width=0.49\linewidth]{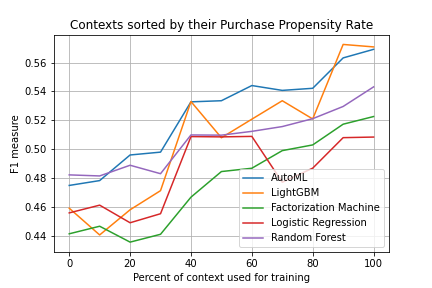}
  \includegraphics[width=0.49\linewidth]{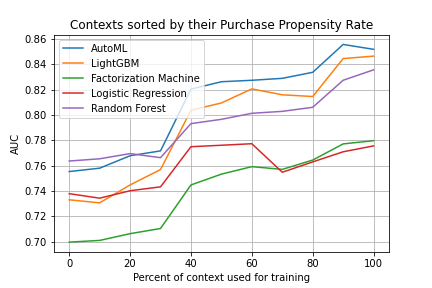}

\caption{Recommendation performance of models trained using different percent of contexts from the head of \textit{Purchase Propensity Frequency} and \textit{Purchase Propensity Rate} distributions for \textit{Leasing} Service.} \label{fig:leasing}
\end{figure}


\end{document}